\documentclass[11pt]{article}

\usepackage {amssymb,latexsym}
\usepackage {amsmath,amsfonts}

\begin{document}


\thispagestyle{empty}

\begin{center}
   {\Large\bf
      Zero Energy States for $SU(N)$: \\
     A Simple Exercise in Group Theory ?
    }

\vspace{3cm}

              M.~Bordemann \\
           Fakult\"at f\"ur Physik, Uni Freiburg \\
           e-mail: mbor@majestix.physik.uni-freiburg.de \\[3mm]

              J.~Hoppe \\
           Max-Planck-Institut f\"ur Gravitationsphysik \\
           e-mail: hoppe@aei-potsdam.mpg.de  \\[3mm]

              R.~Suter \\
           Mathematik Departement ETH Z\"urich \\
           e-mail: suter@math.ethz.ch  \\

\vspace{3cm}

\begin{minipage}{8cm}
 {\small \begin{center} {\bf Abstract} \end{center} 
 We show that the requirement of $S_3\times Spin(d)$ invariance for an 
  ``asymptotically free'' $SU(3)$-Cartan subalgebra wave-function does 
 {\em not} give a unique candidate for a $SU(3)$-invariant zero energy state
 of the $d=9$ supersymmetric matrix model-- nor does it rule out the 
 existence of such a state in the case $d=3$. For $d=9$ we explicitly construct
 various $S_3\times Spin(9)$-invariant wave-functions.}
\end{minipage}        

\end{center}

\newpage


Due to coherent evidence\footnote{See e.g. \cite{KS} and references therein
 (note also \cite{K}); a complete derivation for $SU(2)$ can be found in
 \cite{FGHHY}.}
that, asymptotically, a zero energy state of a $SU(N)$-invariant supersymmetric
matrix model in $d=3,~5$ or $9$ dimensions factorizes into a part 
involving only the Cartan subalgebra degrees of freedom (with effectively
free dynamics) and a part forming supersymmetric harmonic oscillators,
the following ``wishful conjecture'' 
appeared:

\vspace{3mm}

\noindent {\bf CW}: 
For each $N\geq 2$, and $d=9$, the free Laplacian (times the unit operator
in a $2^{(N-1)(d-1)}$-dimensional Fock space) admits exactly one 
$Spin(d)\times S_{N}$ invariant wavefunction (where $S_N$ denotes the
permutation group of $N$ letters) which is square integrable at $\infty$
and harmonic everywhere except at the origin, --whereas for $d=5$ and $3$
no such function exists.

\vspace{5mm}

If true, {\bf CW} would provide further evidence for the global existence of 
a unique
(normalizable) zero energy state for $d=9$, and (assuming effectively free 
dynamics for the Cartan subalgebra part) prove the nonexistence of a global
zero energy state for $d=3$ and $5$.

For $N=2$, the $d=9$ part of {\bf CW} is implied by \cite{HS}, \cite{GH},
while the nonexistence for $d=3$ and $5$ was proven in \cite{HY}, \cite{FGHHY}.

In this note we shall prove {\bf CW} to be {\em wrong}
(for $N=3$), by explicit construction of five linearly independent 
$S_3\times Spin(9)$-invariant,
 and one $S_3\times Spin(3)$-invariant, wave-functions.

\vspace{1.5cm}

For $N=3$, the ingredients are the tensor product of two copies of a 
$2^{d-1}$-dimensional Fock space $\cal H$, each decomposing as
\begin{eqnarray}
  {\cal H}_4        & = &  \{1\} \oplus  \{1\} \oplus  \{2\}    \\
  {\cal H}_{16}     & = &  \{1\} \oplus  \{1\} \oplus   \{1\} \oplus 
                            \{5\} \oplus \{4\} \oplus \{4\} \\
  {\cal H}_{256}    & = &  \{44\} \oplus \{84\} \oplus \{128\}   
                                                   \label{BigFermions}
\end{eqnarray}
under $Spin(d)$, respectively (where we write $\{n\}$ for an
irreducible representation space of dimension $n$), and the space of all 
harmonic polynomials
in $2d$ variables $z=(x,y)$ where $z_a=:x_a$ for $a\leq d$ and $z_a=:y_{a-d}$
for $a>d$. An arbitrary harmonic polynomial of degree $l$ takes the form
\begin{equation}
 h(z) = \sum_{1\leq a_1\cdots a_l\leq 2d} 
               c_{a_1\cdots a_l} z_{a_1}\cdots z_{a_l}
\end{equation}
where the tensor $c$ is totally symmetric and traceless between any two 
indices. As (setting $r:=\sqrt{\sum_a (z_a)^2}$)
\begin{equation}
 \big( \big(\frac{\partial}{\partial r}\big)^2 
        + \frac{2d-1}{r}\frac{\partial}{\partial r}
        - \frac{l(l+2d-2}{r^2} \big) r^{-l-(2d-2)}    =    0, 
\end{equation}
each homogeneous harmonic polynomial of degree $l$ may be multiplied by
$r^{-2l-(2d-2)}$ (to ensure a square integrable fall-off at $\infty$),
without losing harmonicity away from the origin.

The Weyl group for $SU(N)$ is known to be the symmetric group $S_N$,--
which for $N=3$ has only three pairwise inequivalent irreducible
representations, namely the trivial representation in a one-dimensional
module (denoted by $\{1\}$), the alternating representation by 
the sign of the
permutations in a one-dimensional module (denoted by $\epsilon$), and the 
standard representation
in a two dimensional module (denoted by $\rho$) which in some orthonormal basis
takes the following form:
\begin{equation}
 \begin{array}{ccc}
  1 = \left( \begin{array}{cc}
               1 & 0 \\
               0 & 1
             \end{array} \right), 
              &  C  = \frac{1}{2} \left( \begin{array}{cc}
                                  -1    & \sqrt{3} \\
                               -\sqrt{3} &    -1
                                \end{array} \right),
              & C^2 = \frac{1}{2} \left( \begin{array}{cc}
                                  -1    & -\sqrt{3} \\
                                \sqrt{3} &    -1
                                \end{array} \right), \\
  P = \left( \begin{array}{cc}
               1 &  0 \\
               0 & -1
             \end{array} \right), 
              &  P'  = \frac{1}{2} \left( \begin{array}{cc}
                                  -1     & \sqrt{3} \\
                                \sqrt{3} &    1
                                \end{array} \right),
              &  P'' = \frac{1}{2} \left( \begin{array}{cc}
                                   -1     & -\sqrt{3} \\
                                -\sqrt{3} &     1
                                \end{array} \right),
 \end{array}
\end{equation}
where $C$ and $C^2$ denote the two nontrivial cyclic permutations and
$P,P',P''$ denote the three transpositions. Taking traces, the characters
of the above irreducible representations are easily computed to take the
following values on the three conjugacy classes $\{1\}$, $\{C,C^2\}$, and
$\{P,P',P''\}$:
\begin{equation} \label{charactertable}
 \begin{array}{lcccc}
  {\rm for~}    1    & : & 1, &  1, &  1 \\
  {\rm for~}\epsilon & : & 1, &  1, & -1 \\
  {\rm for~}\rho     & : & 2, & -1, &  0 .
 \end{array}
\end{equation}
In the sequel we shall identify the characters with the equivalence class
of irreducible representations they define.

The generators of (the Lie algebra of) $Spin(d)$ can be represented in 
fermionic Fock space $\wedge \mathbb C^{(d-1)(N-1)}$ as operators in
the following way where $s_d:=4,8,16$ for $d=3,5,9$):
\begin{equation}
 \begin{array}{ccc}
M_{d,d-1} & = & \frac{i}{2}\big(
                    \mu_{\alpha n}\partial_{\mu_{\alpha n}}  
                    - \frac{N-1}{4}s_d           \big)  \\
M_{d,j}   & = & \frac{1}{4}{\Gamma^j}_{\alpha\beta}\big(
                   \mu_{\alpha n} \mu_{\beta n}
                    -\partial_{\mu_{\alpha n}}\partial_{\mu_{\beta n}}
                    \big) \\
  M_{d-1,j} & = & \frac{-i}{4}{\Gamma^j}_{\alpha\beta}\big(
                    \mu_{\alpha n} \mu_{\beta n}
                    +\partial_{\mu_{\alpha n}}\partial_{\mu_{\beta n}}
                     \big) \\
  M_{jk}    & = & \frac{1}{2}{\Gamma^{jk}}_{\alpha\beta}\mu_{\alpha n}
                                          \partial_{\mu_{\beta n}}
\end{array}
\end{equation}
Here the $\mu_{\alpha m}$ and $\partial_{\mu_{\beta n}}$ 
($1\leq \alpha, \beta\leq d-1$, $1\leq m,n \leq N-1$) are fermionic
creation (left exterior multiplication) and annihilation (interior product)
operators, $1 \leq j,k \leq d-2$, the $(d-2)$ $\Gamma^j$ are purely 
imaginary, antisymmetric matrices satisfying the anticommutation relations
\[
     \{ \Gamma^j,\Gamma^k \} = 2\delta^{jk} 1_{(d-1)\times (d-1)}
\]
and the $\Gamma^{jk}_{\alpha\beta}$ are the commutators
$\frac{1}{2}[\Gamma^j,\Gamma^k ]_{\alpha\beta}$. For $N=3$ we denote
$\mu_{\alpha 1}$ by $\lambda_\alpha$ and $\mu_{\alpha 2}$ by $\mu_\alpha$.

\vspace{1cm}

For $d=3$, $\mathcal H_4\otimes \mathcal H_4$ gives five $Spin(3)$-singlets,
four $2$-dimensional $Spin(3)$-representations, and one $3$-dimensional one
spanned by
\begin{equation}
   1, \lambda_1\lambda_2+\mu_1\mu_2,\lambda_1\lambda_2\mu_1\mu_2. 
\end{equation}
While the one-dimensional spaces spanned by $\lambda_1\mu_1$, $\lambda_2\mu_2$, and 
$\lambda_1\mu_2+\lambda_2\mu_1$, respectively, are isomorphic to $\epsilon$
under $S_3$, the 
remaining two $Spin(3)$-singlets span a $S_3$-module isomorphic to $\rho$
with basis
\begin{equation}
 |1\rangle:=\frac{1}{\sqrt{2}}(\mu_1\lambda_2-\mu_2\lambda_1),~~
    |2\rangle:=\frac{1}{\sqrt{2}}(\lambda_1\lambda_2-\mu_1\mu_2).
\end{equation}
On the other hand one can easily check that the two $SO(3)$-scalars
$\vec{x}^2-\vec{y}^2$ and $2\vec{x}\cdot\vec{y}$ transform the same way
under the two transpositions $P$ and $P'$ (hence, under $S_3$) as
$|1\rangle$ and $|2\rangle$. Therefore
\begin{equation}
  \psi_{d=3}(\vec{x},\vec{y}) :=
        \frac{1}{r^8}\left(
                (\vec{x}^2-\vec{y}^2)(\lambda_1\lambda_2-\mu_1\mu_2)
               +2\vec{x}\cdot\vec{y}(\mu_1\lambda_2-\mu_2\lambda_1)
             \right)
\end{equation}
is invariant under $Spin(3)\times S_3$, as well as asymptotically normalizable
\[
     \int_{\Lambda}^\infty r^5drr^{-12} ~~ < ~~ \infty
\]
and harmonic
\begin{equation}
   \Delta_{\mathbb R^6}\psi_{d=3}=0,
\end{equation}
thus disproving the $d=3$ part of {\bf CW}.


\vspace{1cm}

For $d=9$, things are somewhat more complicated: while the decompositions
into irreducible $Spin(9)$-modules in the tensor product of two copies of
the fermionic Fock space can be computed using the formulas
(where we agree upon writing $m\{n\}$ for the direct sum of $m$ copies
of the irreducible module $\{n\}$)\footnote{In this note we shall not 
encounter inequivalent irreducible $Spin(d)$-modules of the same dimension.}. 
\begin{eqnarray}
  \{44\}\otimes \{44\}   & = & \{1\} \oplus \{36\} \oplus \{44\} \oplus 
                              \{450\} \oplus \{495\} \oplus \{910\} 
                                       \label{DieKleine}\\
  \{44\}\otimes \{84\}   & = & \{84\} \oplus \{231\} \oplus \{924\} \oplus 
                                          \{2457\}       \\
  \{84\}\otimes \{84\}   & = & \{1\} \oplus \{36\} \oplus \{44\} \oplus 
                        \{84\} \oplus 
                        \{126\} \oplus
                        \{495\} \oplus \{594\} \nonumber \\
                     &   &  \oplus \{924\} \oplus \{1980\} \oplus \{2772\}  
                              \label{DieMittlere}\\
  \{44\}\otimes \{128\}  & = & \{16\} \oplus \{128\} \oplus \{432\} \oplus 
                            \{576\} 
                            \oplus \{1920\} \oplus \{2560\} \nonumber\\
                         &   &                                      \\
  \{84\}\otimes \{128\}  & = & \{16\} \oplus 2\{128\} \oplus 2\{432\} 
                              \oplus \{576\} \oplus \{672\} \oplus 
                                                        \{768\}   \nonumber\\
                     &   & ~~\oplus \{2560\} \oplus \{5040\} \\
 \{128\}\otimes \{128\}  & = & \{1\} \oplus \{9\}  \oplus 2\{36\} \oplus \{44\}
                            \oplus 2\{84\} \oplus 
                                 2\{126\} \nonumber\\
                 &   & \oplus \{156\} \oplus 2\{231\} \oplus \{495\} \oplus 
                        2\{594\}
                        \oplus \{910\} \nonumber\\  
                 &   & \oplus 2\{924\} \oplus \{1650\} \oplus \{2457\} 
                       \oplus \{2772\} \oplus \{3900\} \label{DieNeun}
\end{eqnarray}
found in the literature (see e.g. \cite[p.103, Table 40]{Sla}) the individual
transformation properties under $S_3$ are more difficult to obtain: according
to the general theory of compact Lie groups this 
would amount to a finer decomposition of the above space into tensor 
products of irreducible $S_3$-modules with irreducible $Spin(9)$-modules;
note however that there is only one (!) $9$-dimensional representation on the
right hand sides of the previous six equations (which, therefore, must
be equivalent either to $\{1\}\otimes \{9\}$ or to 
$\epsilon\otimes \{9\}$).


\vspace{0.5cm}

Calculating the decomposition into irreducible
$S_3\times Spin(9)$-submodules of the space of all cubic harmonic
polynomials one finds
\begin{eqnarray}
   Sym^3_{\rm harm} & = & \{1\}\otimes \{9\} \oplus \epsilon\otimes \{9\} 
                           \oplus\rho\otimes \{9\}
                                    \oplus\{1\}\otimes \{156\} \nonumber \\
                               &   & \oplus\epsilon\otimes\{156\}\oplus
                                     \rho\otimes \{156\}\oplus
                                     \rho\otimes \{231\}.
                                           \label{HarmCubic}
\end{eqnarray}
To obtain this, we first used the decomposition into irreducible 
$Spin(9)$-submodules
\begin{eqnarray}
  Sym^3(\mathbb R^9 \oplus \mathbb R^9) & = &
                     2\{Sym^3(\mathbb R^9)\}\oplus 
                        2\{Sym^2(\mathbb R^9)\otimes \mathbb R^9\} 
                                                     \nonumber \\
                    & = &  2\{156\oplus 9\} \oplus 
                                   2\{\{1\oplus 44\}\otimes 9\} 
                                                     \nonumber \\
                    & = &  6\{9\} \oplus 4\{156\} \oplus 
                              2\{231\} \label{SpinKub} \\
\end{eqnarray}
as well as the decomposition into irreducible $S_3$-submodules 
(using $\rho\otimes \rho= \{1\}\oplus \epsilon \oplus \rho$, see the 
character table (\ref{charactertable})):
\begin{eqnarray}
Sym^3(\mathbb R^9 \oplus \mathbb R^9)
                      & = & 165\{1\} \oplus 165\epsilon \oplus 405\rho
\end{eqnarray}
where these have to be distributed among (\ref{SpinKub}) (which turns out to be
doable by just counting dimensions); finally, one of the 
emerging $\rho\otimes \{9\}$ modules had to be dropped, as it 
corresponds to the submodule spanned by the $r^2x_a$ and $r^2y_a$ 
($1\leq a\leq 9$) which does not contain any nonzero harmonic functions.
In any case, as (\ref{HarmCubic}) contains both $\{1\}\otimes \{9\}$ and
$\epsilon\otimes \{9\}$, the single $Spin(9)$-submodule of dimension $9$
in the fermionic sector (\ref{DieNeun}) --no matter whether it is of type 
$\{1\}\otimes \{9\}$ or $\epsilon\otimes \{9\}$--can be matched in the final
tensor products of bosons and fermions to form a $S_3\times Spin(9)$ invariant
wave function of the form
\begin{equation} \label{CubicGroundstate}
    \Psi(z) = r^{-22}\sum_{s=1}^9\psi_s(z)|9;s\rangle
\end{equation}
where the $|9;s\rangle$ form an orthonormal basis of the 
$\{9\}$-module in (\ref{DieNeun}) and the $\psi_s$ form an 
orthonormal basis of the module of the same type in the space of harmonic 
cubic polynomials.


\vspace{0.5cm}

One could be tempted to deduce that this must be the Cartan subalgebra
factor in the asymptotic form of {\em the} $d=9$, $N=3$ zero energy 
wave-function.
However, at least (!) four other $S_3\times Spin(9)$ candidate wave-functions
exist (thus disproving the uniqueness of the $d=9$ part of {\bf CW});
to see this, replace $9$ by $156$ in the previous argument (which gives the
second invariant ground state), or consider the space of all quartic 
harmonic polynomials, which after some work can be concluded to 
decompose into
\begin{eqnarray}
  Sym^4_{\rm harm}(\mathbb R^{18}) & = & 
                            \{1\}\otimes \{450\}~\oplus~ 2\rho\otimes \{450\}
                                 ~ \oplus~\rho\otimes\{910\}~
                                    \oplus~\epsilon\otimes\{910\}\nonumber\\
                            &   & ~\oplus~\{1\}\otimes\{495\}~\oplus~
                                   \{1\}\otimes \{44\}~\oplus~ 
                                   \epsilon\otimes\{44\}
                                   ~\oplus~2\rho\otimes\{44\}\nonumber\\
                            &   &  \oplus~
                                    \rho\otimes\{36\}~\oplus~\rho\otimes\{1\}~
                                    \oplus~\{1\}\otimes \{1\} \label{HarmQuart}
\end{eqnarray}
under $S_3\times Spin(9)$.

Now note that the r.h.s~of eqs (\ref{DieKleine})-(\ref{DieNeun}) contains
three $Spin(9)$-submodules of dimension $495$. The submodule of 
dimension $128$ in (\ref{BigFermions}) consists of all forms of odd degree
over $\mathbb R^8$ (see e.g. \cite{GH}) whereas the two submodules of dimension
$44$ and $84$ give all corresponding forms of even degree. It follows that
the transposition $P\in S_3$, which maps $(\lambda_\alpha,\mu_\alpha)$ to 
$(\lambda_\alpha,-\mu_\alpha)$,
is equal to the identity on the two irreducible modules of dimension $495$ 
contained in (\ref{DieKleine}) and (\ref{DieMittlere}) since they imply even 
forms in $\lambda$ and $\mu$ separately, whereas it is equal to minus identity
on the irreducible module of dimension $495$ contained in (\ref{DieNeun}) since
it consists of forms which are odd both in $\lambda$ and in $\mu$. Out of the
$6$ apriori possibilities to decompose the sum of the three modules of
dimension $495$ into irreducibles under $S_3\times Spin(9)$, the two {\em not}
containing $\{1\}\otimes \{495\}$, namely 
$\epsilon\otimes \{495\}~\oplus~\rho\otimes \{495\}$ and $3\epsilon\otimes\{495\}$,
are therefore impossible, as their $+1$-eigenspace under $P$ would be too 
small. Therefore the fermionic sector contains at least one $S_3\times
Spin(9)$-submodule of type $\{1\}\otimes \{495\}$ that can be matched with the
irreducible submodule of dimension $495$ in (\ref{HarmQuart}) in the same
manner as in the cubic case (see (\ref{CubicGroundstate})), viz.
\begin{equation} \label{QuarticGroundstate}
    \Psi(z) = r^{-24}\sum_{s=1}^{495}\psi_s(z) |495;s\rangle
\end{equation}
where the $|495;s\rangle$ form an orthonormal basis of the 
$\{1\}\otimes \{495\}$-module in the fermionic sector and the $\psi_s$ form an 
orthonormal basis of the module of the same type in the space of harmonic 
quartic polynomials. The $\psi_s$ can all be chosen out of the quartic
polynomials taking the following 
form\footnote{Note that the Weyl-invariant $450$-dimensional representation
$\bar{\psi}^{(450)}$ occurring in (\ref{HarmQuart}) is of the form
$c_{stuv}( x_sx_tx_ux_v+y_sy_ty_uy_v+2x_sx_ty_uy_v)$ with $c_{stuv}$ totally symmetric
and traceless; contracting it with the unique, Weyl-invariant, $450$-dimensional
representation formed out of the fermions, yields yet another 
$S_3\times Spin(9)$-invariant wavefunction.}:
\begin{equation}
 \psi_s(x,y) = \sum {c^{(s)}}_{a_1a_2a_3a_4}(x_{a_1}x_{a_2}-
                                \frac{1}{9}x_ax_a\delta_{a_1a_2})
                             (y_{a_3}y_{a_4}-\frac{1}{9}y_ay_a\delta_{a_3a_4})
\end{equation}
where $c^{(s)}$ is a tracless tensor of rank four in $\mathbb R^9$, corresponding
to the Young diagram $\boxplus$.
This will give a third candidate for the Cartan 
subalgebra factor of an asymptotic zero-energy state of the $d=9$ $SU(3)$
matrix model.

A fourth candidate is obtained by noting that the decomposition
(\ref{HarmQuart}) contains the three pairwise nonequivalent irreducible 
$S_3\times Spin(9)$-modules $\{1\}\otimes \{44\}$, $\epsilon\otimes \{44\}$, 
and $\rho\otimes \{44\}$. According to eqs (\ref{DieKleine}), 
(\ref{DieMittlere}), and (\ref{DieNeun}) the
fermionic sector contains two of these modules which again match to at least
two $S_3\times Spin(9)$-invariant states of the form
\begin{equation} \label{QuarticGroundstateII}
    \Psi(z) = r^{-24}\sum_{s=1}^{44}\psi_s(z)|44;s\rangle
\end{equation}
with the obvious notations.

Last, but not the least, consider the space of all quadratic 
harmonic polynomials which decomposes as
\begin{equation}
   Sym^2_{\rm harm}= \rho\otimes\{1\} \oplus \rho\otimes\{44\} \oplus
                      \{1\}\otimes\{44\} \oplus \epsilon\otimes\{36\}
\end{equation}
under $S_3\times Spin(9)$ (corresponding to the space spanned by
$\vec{x}^2-\vec{y}^2$ and $2\vec{x}\cdot\vec{y}$, the space of polynomials
$\sum c_{ab}(x_ax_b-y_ay_b)$ and $\sum c_{ab}(x_ay_b)$ with $c$ symmetric
traceless, the space of polynomials $\sum c_{ab}(x_ax_b+y_ay_b)$ with $c$ 
symmetric traceless, and the space of polynomials $\sum d_{ab}(x_ay_b)$
with $d$ antisymmetric, respectively). Again, one of the above modules
containing $\{44\}$ as a factor can be matched with an equivalent fermionic 
one, as at  least one of the two fermionic modules containing $\{44\}$ as
a factor is {\em not} isomorphic to $\epsilon\otimes\{44\}$: otherwise
all the three $Spin(9)$-modules of type $\{44\}$ in (\ref{DieKleine}),
(\ref{DieMittlere}), and (\ref{DieNeun}) would have to change sign
under the transposition $P$ which is not the case for the two modules
of type $\{44\}$ in (\ref{DieKleine}) and (\ref{DieMittlere}),
compare the discussion of $\{495\}$.

\vspace{0.5cm}

Finally, note that
while $H\psi=0$ implies $Q\psi=0$ if $Q$ is hermitean and
$H=Q^2$, care is needed when looking at the corresponding differential
equations only asymptotically, or when deriving `effective' operators
(as done in \cite{KS}). As stressed by A.Smilga \cite{KS,Smi}, it may well
be that the additional condition $Q_{\rm eff}\psi=0$ will single out
a unique $d=9$ wave-function and/or exclude any $d=3,5$ 
$S_N\times Spin(d)$-wave-function. It should be easy to test this
using our $S_3\times Spin(d)$-invariant wave-functions.

\vspace{0.5cm}

\noindent {\large\bf Acknowledgements}: J.H. would like to thank J.Plefka and 
A.Smilga for valuable discussions, while M.B. thanks the Albert-Einstein-Institut 
for friendly hospitality.

\vspace{0.5cm}

\noindent {\large \bf Note added}: It is easy to see that neither
(\ref{QuarticGroundstateII}$)_{\{1\}\otimes\{44\}}$
(nor (\ref{QuarticGroundstate})), nor $r^{-24}$ times the 
$\{1\}\otimes \{1\}$ in (\ref{HarmQuart}) times
a Weyl invariant singlet formed
out of the fermions, cf. (\ref{DieKleine})-(\ref{DieNeun}), nor 
$r^{-24}\bar{\psi}^{(450)}_{stuv}|stuv\rangle$ can be supersymmetric: 
just note that each of the free supercharges
is a sum of two terms, the first of which  changes the number of $\lambda$'s
by $\pm 1$, and squares to the $x$-Laplacian, while the second, squaring to
the $y$-Laplacian, does not contain $\lambda$ nor $\partial_\lambda$.

\end{document}